\documentclass[final,5p,times,twocolumn]{elsarticle} 
\usepackage{lineno,hyperref}
\modulolinenumbers[5]

\journal{NIM A}

\begin{document}

\begin{frontmatter}

\title{	Particle Identification with the ALICE Transition Radiation Detector}

\author{Yvonne Pachmayer\fnref{myfootnote}}
\ead{pachmay@physi.uni-heidelberg.de}
\address[myfootnote]{Physikalisches Institut, University of Heidelberg (Germany)}
\author{for the ALICE Collaboration}




\begin{abstract}
The Transition Radiation Detector (TRD) provides particle identification in the ALICE central barrel. In particular, it allows electron identification via the measurement of transition radiation for $\rm p >$ 1 GeV/$c$, where pions can no longer be rejected sufficiently via specific energy loss in the ALICE Time Projection Chamber. The ALICE TRD is uniquely designed to record the time evolution of the signal, which allows even better electron/pion separation. In addition, the electron identification capability of the TRD can be used on-line to trigger at level 1. The particle identification and its performance in pp, p-Pb and Pb--Pb collisions employing various methods, such as truncated mean signal, one- and two-dimensional likelihood on integrated charge and neural network, will be presented. The measurement of J/$\psi$ mesons in Pb--Pb collisions is given as a case study to show how well the TRD contributes to physics analyses due to its excellent pion suppression. 
\end{abstract}


\end{frontmatter}

\linenumbers

\section{Introduction}\label{Section_Intro}

The ALICE Transition Radiation Detector (TRD) \cite{ALICETRD} performs electron identification and triggers at level 1 on identified particles. It thus allows semi-leptonic decays of heavy-flavour hadrons, di-electron mass spectra of heavy quarkonia states and jets to be studied \cite{YPachmayer}. These observables are important probes of the quark-gluon plasma created in heavy-ion collisions at the LHC. For reference purposes the measurements are also performed in pp and p-Pb collisions. 

The ALICE TRD consists of 522 chambers arranged in 6 layers at a radial distance \textit{r} (2.9 $\leq$ \textit{r} \mbox{$\leq$ 3.7 m)} from the beam axis. Each chamber comprises a radiator and a gas detector with a 3 cm drift region and a 0.7 cm multi-wire proportional chamber (MWPC) with amplification on anode wires. The signal induced on the segmented cathode plane is read out and processed in a custom-built charge-sensitive preamplifier-shaper circuit and digitized by a 10 MHz ADC to sample the temporal evolution of the signal. The induced charges are sampled every \mbox{100 ns.} The front-end electronics are directly mounted on the back of each TRD chamber. 

\begin{figure}[htb]
   \centering
\includegraphics[width=.37\textwidth]{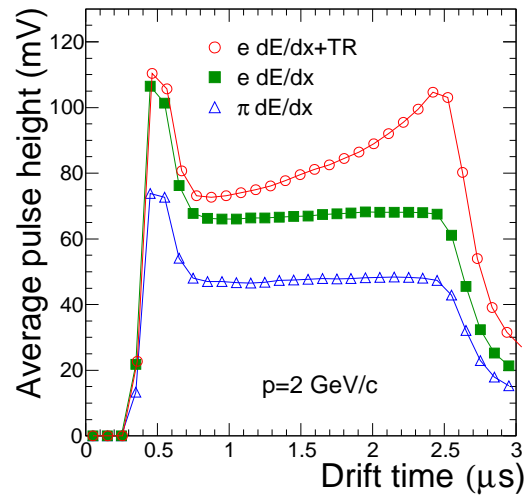} 
  \caption[]{Average pulse height vs drift time for
pions and electrons (with and without radiator). The results were obtained from testbeam measurements \cite{Anton}.}
   \label{Figure_ph}
\end{figure}

Figure~\ref{Figure_ph} shows the respective recorded time evolution of the signal for electrons and pions with momenta of 2 GeV/$c$. The peak at early times originates from the amplification region and the plateau at intermediate times is due to the drift region. The average pulse height is higher for electrons, because in this momentum region the specific energy loss of electrons is larger than that of pions.

When relativistic electrons ($\gamma \gtrsim 800$) travel through the radiator, crossing the many boundaries between media with different dielectric constants, on average one Transition Radiation (TR) photon with an energy in the X-ray range is created per traversing electron. The photons are detected in the Xe/CO$_2$ filled TRD chambers \citep{RBailhache}, where they deposit their energy on top of the ionization signals from the particle track. This contribution is visible in Fig.~\ref{Figure_ph} at late drift times, because the absorption of the TR photon happens preferentially close to the radiator. 

\begin{figure}[htb]
   \centering
\includegraphics[width=.49\textwidth]{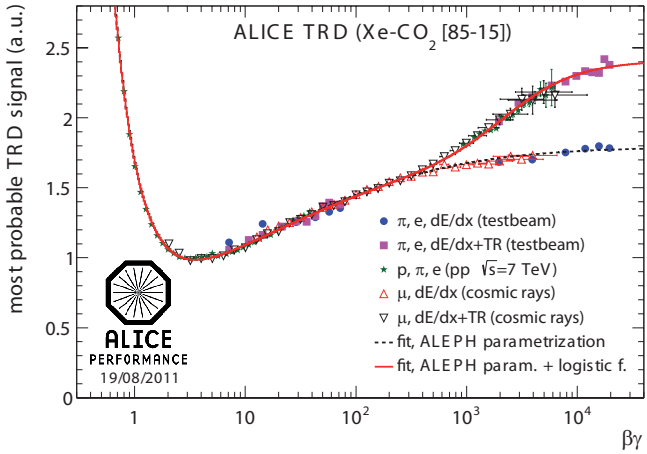} 
  \caption[]{Most probable TRD signals from testbeam runs, pp collisions at \mbox{$\rm \sqrt{s} = 7 ~TeV$} and cosmic ray measurements \cite{XLu}.}
   \label{Figure_mpv}
\end{figure}

Specific energy loss and TR measurements were performed in the $\beta\gamma$ range $\rm 1-10^{4}$ with (i) pions and electrons from testbeam runs at CERN PS in 2004 \cite{RBailhache}, (ii) protons, pions and electrons in pp collisions at $\rm \sqrt{s} = 7~TeV$  \cite{MFasel} and (iii) muons detected in ALICE cosmic runs \cite{XLu}. The latter experiment permits the measurement of both dE/dx + TR and dE/dx only signals in the TRD by selecting cosmic tracks entering the TRD chambers from the radiator side or the opposite side, respectively. The dependence of the most probable signal on $\beta\gamma$  is shown in Fig. \ref{Figure_mpv}. The onset of TR from cosmic muons is distinctly visible and the TR signal from TeV muons is consistent with that from GeV electrons in the other measurements. This allows reference distributions for particle identification with the TRD to be determined over a broad momentum range.

\hspace{-0.2cm}
\section{Particle Identification}\label{Section_PID}

The TRD provides particle identification on a track-by-track basis. Several methods are in use:
\begin{itemize}
\item Truncated mean signal,
\item One-dimensional likelihood,
\item Two-dimensional likelihood and
\item Neural network.
\end{itemize}

\begin{figure}[htb]
   \centering
\includegraphics[width=.49\textwidth]{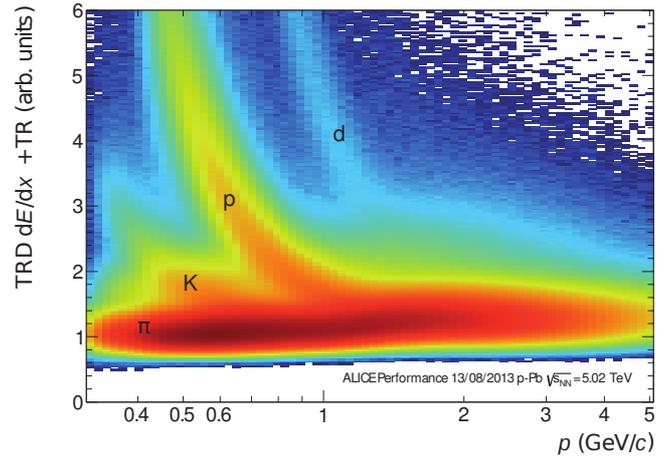} 
  \caption[]{Truncated mean signals as a function of momentum for charged particles in minimum bias data from p-Pb collisions at
 $\rm \sqrt{s} = 5.02 ~TeV$ \cite{XLu2}.}
   \label{Figure_trunc}
\end{figure}

The truncated mean signal, the combined signal of specific ionization and transition radiation, is shown versus momentum in Fig.~\ref{Figure_trunc} for p-Pb collisions at $\rm \sqrt{s} = 5.02 ~TeV$ \cite{XLu2}. As can be seen, this method allows identification of hadrons and light nuclei. \newline

\begin{figure}[htb]
   \centering
\includegraphics[width=.5\textwidth]{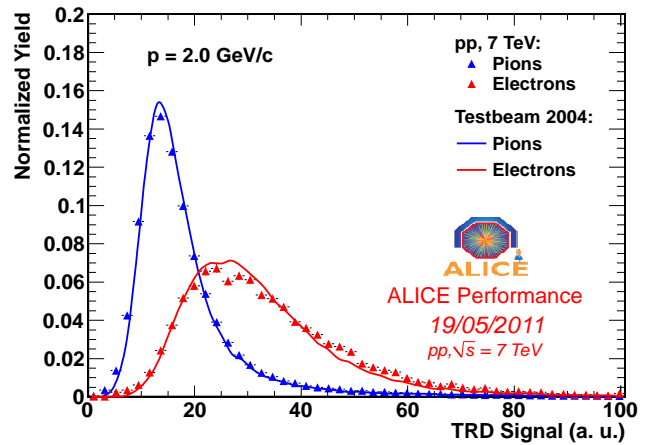} 
  \caption[]{Total integrated charge measured in a single TRD chamber for electrons and pions in pp collisions, in comparison with results from testbeam measurements. In pp collisions, the electrons and pions were selected via topological cuts from photon conversions and $\rm K^{0}_{s}$ decays, respectively.}
   \label{Figure_trackletcharge}
\end{figure}

The simplest dedicated electron identification method in use is the one-dimensional likelihood (LQ1D) on the total integrated charge measured in a single TRD chamber \mbox{(tracklet) \cite{MFasel}.} Figure~\ref{Figure_trackletcharge} shows the total charge measured in a single chamber for electrons and pions in pp collisions at $\rm \sqrt{s} = 7 ~TeV$. Clean samples of electrons and pions were obtained by selecting tracks originating from the decay $\gamma \rightarrow e^+ e^-$ and $K^{0}_{s} \rightarrow \pi^+ \pi^-$, respectively, via topological cuts and particle identification with the Time Projection Chamber (TPC) and the Time-of-Flight detector (TOF) in ALICE.
The average charge deposit of electrons is higher than that of pions, because of the larger specific energy loss and transition radiation. \newline

In testbeam measurements at CERN PS in 2004, charge deposit distributions were recorded for electrons and pions in the momentum range 1 to 10 GeV/$c$ \citep{Anton}. The respective charge deposit distributions describe the results from collision data well (see Fig.~\ref{Figure_trackletcharge}), and can thus be used as references for particle identification. Respective reference distributions for muons, kaons and protons were obtained via parametrizations from GEANT3. 

\begin{figure*}[!bht]
     \hspace{0.2cm}
     \begin{minipage}{0.39\textwidth}
      \centering
       \includegraphics[width=1\textwidth]{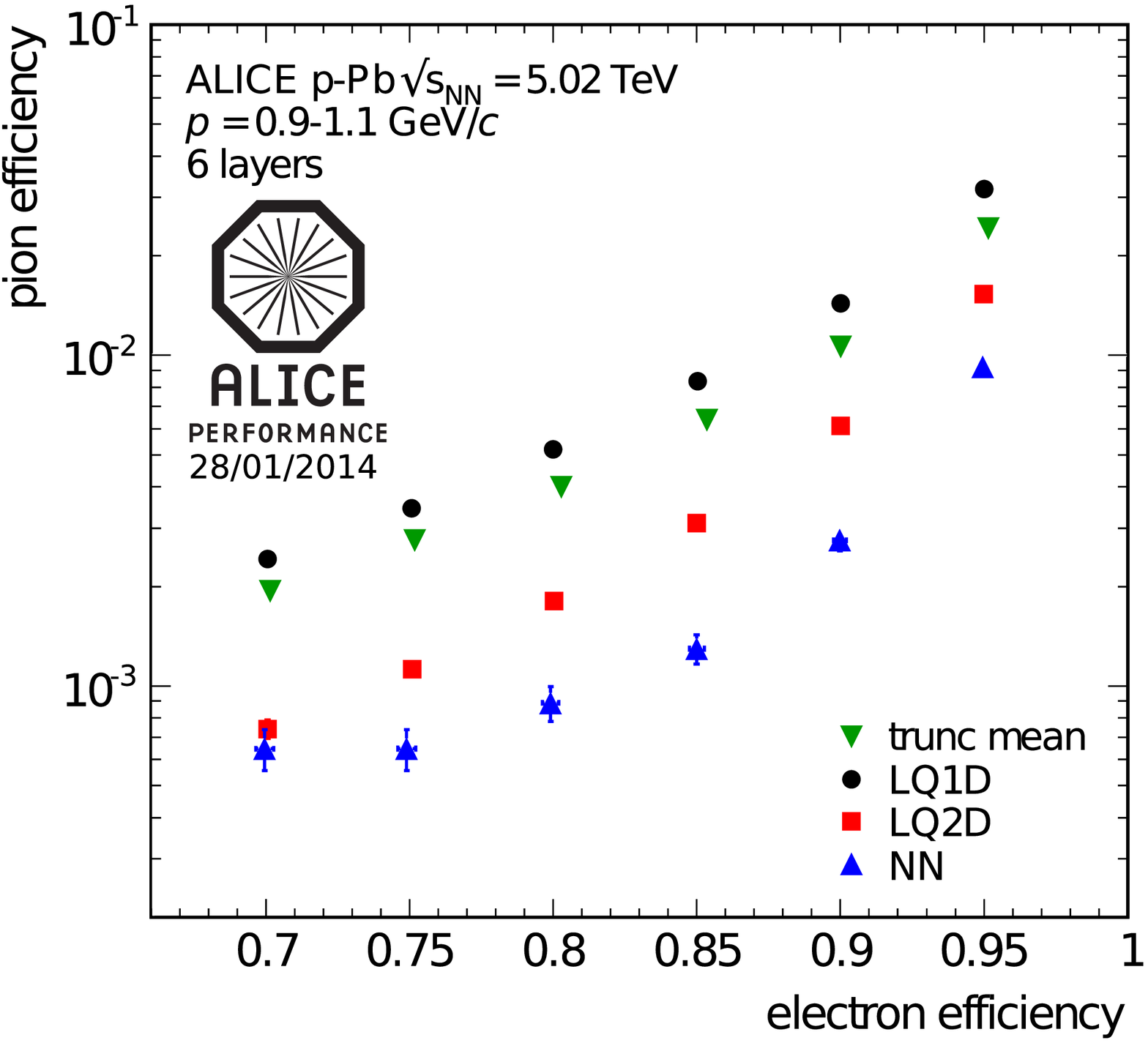}
       \caption{Pion efficiency versus electron efficiency for various particle identification methods.}\label{Figure_pioneffi}
     \end{minipage}
     \hspace{2.2cm}
     \begin{minipage}{0.39\textwidth}
      \centering
     \includegraphics[width=1\textwidth]{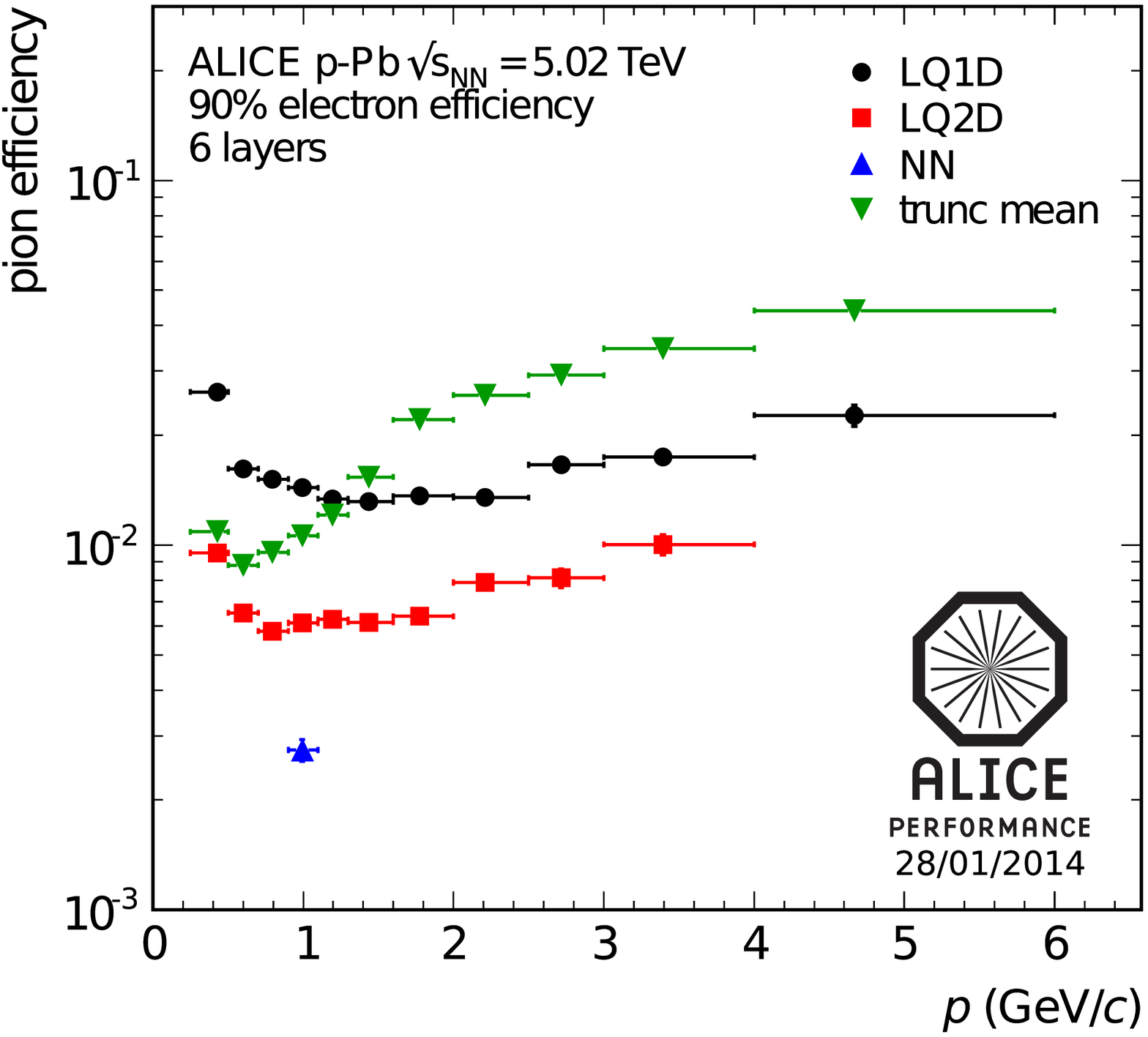} 
       \caption{Momentum dependence of the pion efficiency. The results are for tracks passing 6 layers of the TRD and satisfying an electron efficiency of 90 \%.}\label{Figure_pdependence}
     \end{minipage}
   \end{figure*}

For each particle passing the TRD, likelihood values for electrons, pions, muons, kaons and protons are calculated for each chamber using the reference distributions. Next the information of at least four TRD layers is combined via the Bayesian formula, providing global track particle identification. 

In the analysis, pions, i.e. hadrons, are then rejected in the TRD by applying a momentum-dependent cut on the likelihood value for electrons, providing a specified electron efficiency. The cut was tuned and cross-checked by studying the clean reference sample of electrons from photon conversions. \newline

The two-dimensional likelihood method (LQ2D) \cite{DLohner} and the neural network (NN) \cite{CAdler} make use of the temporal evolution of the TRD signal by splitting the signal, depicted in Fig.~\ref{Figure_ph}, into two slices and into seven slices in time, respectively. The references for the summed charges in both slices as well as the training sample for the neural network were obtained from clean electron and pion samples in collision data. \newline

To quantify the electron/pion separation for the different methods clean samples of electrons and pions are used. The pion efficiency, the fraction of pions incorrectly identified as electrons, is depicted in Fig.~\ref{Figure_pioneffi} as a function of electron efficiency for all described methods. The results at 90 \% electron efficiency confirm the design values obtained in testbeam measurements \cite{Anton}. The LQ2D and NN methods significantly improve the pion suppression (the inverse of pion efficiency) compared with the LQ1D and truncated mean signal method, since these exploit the temporal evolution of the recorded TRD signal. By sacrificing some electron efficiency an even improved pion rejection can be obtained for all methods. \newline
The results shown in Fig.~\ref{Figure_pioneffi} are for tracks passing 6 layers of the TRD. With a decreasing number of tracklets, the pion rejection factor decreases as expected \cite{YPachmayer}.\newline
The momentum dependence of the pion efficiency is shown in Fig.~\ref{Figure_pdependence} for the various electron identification methods for tracks passing 6 layers of the TRD and satisfying an electron efficiency of 90 \%. For the LQ1D and LQ2D methods the pion efficiency decreases initially, because of the onset of TR production (see Fig.~\ref{Figure_mpv}). As expected, the pion suppression gets weaker with further increasing momenta, because the TR production starts to saturate and the pion energy loss increases, making the electron/pion separation less efficient. The LQ2D methods lacks necessary references for momenta above 4 GeV/$c$. The truncated mean signal method shows very good pion rejection at low momenta, where the d$E$/d$x$ measurement dominates. Above, the rejection power decreases, because the TR contribution, yielding higher charge deposits, is more likely to be removed in the truncation.

\section{TRD Electron Identification in Physics Analysis}

\begin{figure}[hbt]
   \centering
\includegraphics[width=.48\textwidth]{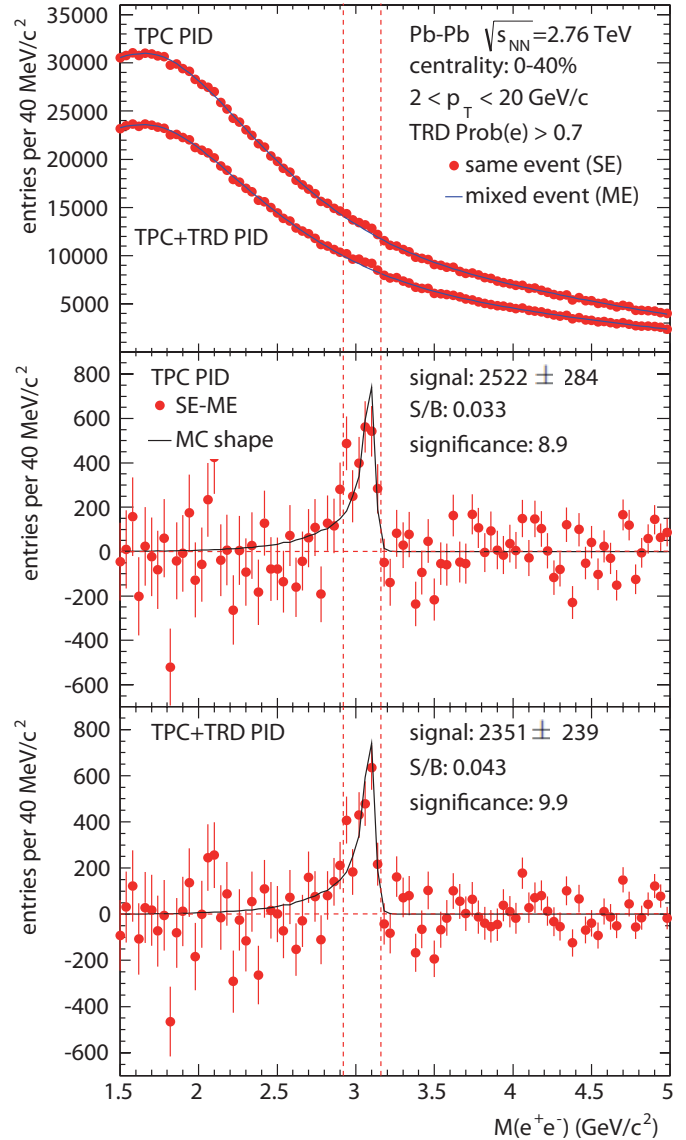} 
  \caption{Invariant mass distribution with TPC-only and TPC+TRD particle identification for 0-40 \% most central Pb--Pb collisions.}
   \label{Figure_jpsi}
\end{figure}

An illustrative case for the application of the TRD electron identification is the reconstruction of the decay of the J/$\psi$ meson into an electron and a positron.
Figure~\ref{Figure_jpsi} shows the invariant mass distribution of $e^+e^-$ pairs in 0-40 \% most central Pb--Pb collisions at $\rm \sqrt{s_{NN}} = 2.76~TeV$, identified by using the TPC only as well as the TPC and TRD combined. 

In both cases, the leptons with momenta $\rm p >$ 2 GeV/$c$ were identified through their specific energy loss in the TPC (inclusion cut $\rm -1.5~\sigma, +3~\sigma$). Furthermore a $\rm \pm 3.5~\sigma$ and $\pm 4~\sigma$ exclusion cut for pions and protons, respectively, was applied. After background subtraction, one finds for the TPC-only case 2552 $\pm$ 284 J/$\psi$ candidates in the invariant mass region \mbox{2.92-3.16 GeV/$c^2$.} The signal-to-background ratio is 0.033 and the significance \mbox{is 8.9.}

In the combined TPC+TRD case, the LQ2D method is additionally applied for electron identification in the TRD by requiring an electron likelihood of at least 0.7 for tracks with at least 4 tracklets. The strong discrimination power of the TRD leads to a clear reduction of combinatorial background in comparison with the TPC-only analysis. 
The signal-to-background ratio increases by 30 \%. 
The remaining combinatorial background is dominated by electrons from photon conversion and semi-leptonic heavy-flavour decays.
In the Pb--Pb run 2011 (data shown), the TRD covered 13/18 of the azimuthal angle. Thus the TRD electron identification was used whenever available to avoid signal loss. Despite the reduced coverage, the results already illustrate the strength of the TRD electron identification.

\section{On-line Electron Identification}

To significantly enrich samples of electrons originating from heavy-flavour decays and quarkonia, the TRD can be used to select events with electrons at trigger level 1 \cite{JKlein}. 

In the front-end electronics, track segments (tracklets) are reconstructed locally and the total integrated charge is transformed into an electron likelihood using a look-up table. This table was created from a clean sample of electrons and pions from collision data (see Section~\ref{Section_PID}). The information of at least 4 layers is combined by averaging over the contributing tracklets. For the trigger thresholds are defined to achieve a specified electron efficiency. In addition, a threshold on the minimum momentum, derived from a linear fit to all tracklets, can be set. Gain variations on a pad-by-pad level are corrected and variations as a function of time due to pressure changes are corrected by high voltage adjustments. For an overview of the TRD trigger see \cite{JKlein}.

The TRD electron trigger was successfully \mbox{operated in} pp collisions at \mbox{$\rm \sqrt{s} = 8 ~TeV$} and p-Pb collisions at \mbox{$\rm \sqrt{s_{NN}}  = 5.02 ~TeV$.} The analysis of the recorded data is ongoing.

\section{Conclusion and Outlook}\label{Section_Summary}

The TRD provides excellent particle identification and allows the enhancement of rare probes due to its trigger capabilities. It thus significantly expands the physics reach in ALICE. 
The two-dimensional likelihood method and the neural network yield a pion suppression factor significantly better than 100 at an electron efficiency of 90 \%, and thus reach the design value already obtained in testbeam measurements.
The pion suppression for the LQ2D method is about a factor two better compared with the one-dimensional likelihood method, because of the usage of the temporal evolution of the TRD signal. The latter method on the other hand is more simple and robust, because it does not depend on the quality of the drift velocity and the time offset calibration.
The application of the TRD particle identification algorithms in further analysis strategies, including both electron and hadron identification methods, and the analysis of TRD triggered data, are ongoing.

\section*{References}

\bibliography{rich_Pachmayer_bibfile}

\begin{thebibliography}{10}
\expandafter\ifx\csname url\endcsname\relax
  \def\url#1{\texttt{#1}}\fi
\expandafter\ifx\csname urlprefix\endcsname\relax\def\urlprefix{URL }\fi
\expandafter\ifx\csname href\endcsname\relax
  \def\href#1#2{#2} \def\path#1{#1}\fi

\bibitem{ALICETRD}
{ALICE C}ollaboration, {A} {T}ransition {R}adiation {D}etector for {E}lectron
  {I}dentification within the {ALICE} {C}entral {D}ector, {CERN/LHCC 99-13,
  LHCC/P3-Addendum} 2, 1999; {ALICE} {T}ransition {R}adiation {D}etector
  {T}echnical {D}esign {R}eport, {ALICE TDR} 9, {CERN/LHCC} 2001-021.

\bibitem{YPachmayer}
{Y.~Pachmayer [ALICE Collaboration]}, Physics with the {ALICE T}ransition
  {R}adiation {D}etector, Nucl. Instrum. Meth A 706 (2013) 6--11.
\newblock \href {http://dx.doi.org/10.1016/j.nima.2012.05.016}
  {\path{doi:10.1016/j.nima.2012.05.016}}.

\bibitem{RBailhache}
{R.~Bailhache and C.~Lippmann [ALICE TRD Collaboration]}, New test beam results
  with prototypes of the {ALICE TRD}, Nucl. Instrum. Meth A 563 (2006)
  310--313.
\newblock \href {http://dx.doi.org/10.1016/j.nima.2006.02.157}
  {\path{doi:10.1016/j.nima.2006.02.157}}.

\bibitem{Anton}
{A.~Andronic et al. [ALICE TRD Collaboration]}, Electron identification
  performance with {ALICE TRD} prototypes, Nucl. Instrum. Meth A 522 (2004)
  40--44.
\newblock \href {http://dx.doi.org/10.1016/j.nima.2004.01.015}
  {\path{doi:10.1016/j.nima.2004.01.015}}.

\bibitem{MFasel}
{M.~Fasel}, Single-electron analysis and open charm cross section in
  proton-proton collisions at $\sqrt{s}$ = 7 {T}e{V}, doctoral thesis, {TU
  D}armstadt, {O}ctober 2012.

\bibitem{XLu}
{X.~Lu [ALICE Collaboration]}, Energy loss signals in the {ALICE TRD}, Nucl.
  Instrum. Meth A 706 (2013) 16--19.
\newblock \href {http://dx.doi.org/10.1016/j.nima.2012.05.015}
  {\path{doi:10.1016/j.nima.2012.05.015}}.

\bibitem{XLu2}
{X.~Lu}, Exploring the performance limits of the {ALICE T}ime {P}rojection
  {C}hamber and the {T}ransition {R}adiation {D}etector for measuring
  identified hadron production at the {LHC}, doctoral thesis, {U}niversitaet
  {H}eidelberg, {G}ermany, {O}ctober 2013.

\bibitem{DLohner}
{D.~Lohner}, Anisotropic flow of direct photons in {Pb}--{Pb} collisions at
  $\sqrt{s_{NN}}$ = 2.76 {T}e{V}, doctoral thesis, {U}niversitaet {H}eidelberg,
  {G}ermany, {O}ctober 2013.

\bibitem{CAdler}
{C.~Adler et al. [ALICE TRD Collaboration]}, Electron/{P}ion {I}dentification
  with {ALICE TRD P}rototypes using a {N}eural {N}etwork {A}lgorithm, Nucl.
  Instrum. Meth. A 552 (2005) 364--371.

\bibitem{JKlein}
{J.~Klein [ALICE Collaboration]}, Triggering with the {ALICE TRD}, Nucl.
  Instrum. Meth A 706 (2013) 23--28.
\newblock \href {http://dx.doi.org/10.1016/j.nima.2012.05.011}
  {\path{doi:10.1016/j.nima.2012.05.011}}.

\end{thebibliography}

\end{document}